\documentclass[3p,times]{elsarticle}
\usepackage{graphicx}
\usepackage{amssymb}
\usepackage{amsmath}
\usepackage{graphics}
\usepackage[T1]{fontenc}
\journal{Annals of Physics (prospectively)}

\begin{document}

\begin{frontmatter}
\title{A source fragmentation approach to interacting quantum field theory}
\author{Peter Morgan}
\address{Physics Department, Yale University, New Haven, CT 06520, USA.}
\ead{peter.w.morgan@yale.edu}
\date{\today}

\begin{abstract}
A corollary to the Reeh-Schlieder theorem is proved: that the time-ordered Vacuum Expectation Values and the S-matrix of a regularized Lagrangian quantum theory can be approximated by a local operator that uses nonlinear functionals of a locally supported source function.
For the Wightman axioms, this suggests a modification that takes the algebra of measurement operators \emph{not} to be generated by an operator-valued distribution.
The use of operator-valued nonlinear functionals of a source function introduces many abstract fragments of the source to give a well-defined top-down construction of interacting quantum fields, in contrast to a bottom-up blocking and scaling construction or to analyzing response to changing renormalization scales.
The construction can also be thought of as solving a localized inverse problem for the interacting dynamics or as a generating function for multi-point bound state fields.
\end{abstract}
\begin{keyword}
Quantum Field Theory, Regularization, Renormalization, Signal Analysis
\end{keyword}
\end{frontmatter}

\newcommand\Half{{\scriptstyle\frac{\scriptstyle 1}{\raisebox{-0.4ex}{$\scriptstyle 2$}} }}
\newcommand\rmd{{\mathrm{d} }}
\newcommand\rme{{\mathrm{e} }}
\newcommand\rmi{{\mathsf{i} }}
\newcommand\Unit{{\hat{\mathbf{1}} }}
\newcommand\kInt[1]{{\frac{\rmd^4#1}{(2\pi)^4}}}

\newcommand\VEV[1]{{\langle0|#1|0\rangle}}
\newcommand\TO[1]{{\mathsf{T}\left[#1\right]}}
\newcommand\TOpv[1]{{\mathsf{T_{\!PV\!}}\left[#1\right]}}

\newcommand\source{{\boldsymbol{j}}}
\newcommand\sourcef{{\boldsymbol{f}}}
\newcommand\sourceg{{\boldsymbol{g}}}
\newcommand\Supp[1]{{\mathsf{Supp}[#1]}}
\newcommand\StarAlgebra{{\raisebox{-0.4ex}{*}-algebra}}

\newtheorem{corollary}{Corollary}

\newenvironment{myquote}[1]%
  {\list{}{\leftmargin=3em\rightmargin=#1%
           \topsep=0ex\partopsep=0ex\parsep=0ex}\item[]}%
  {\endlist}
\newenvironment{bulletpoints}%
  {\list{$\blacktriangleright$}{\leftmargin=3em%
           \topsep=0ex\partopsep=0ex\parsep=-0.3ex}\item[]}%
  {\endlist}


\section{Introduction}
Renormalization is a major difficulty in our understanding of interacting quantum fields, in part because it creates an apparently unbridgeable gap between physicists' practice and mathematically axiomatic approaches.
For the Wightman axioms and for the Haag-Kastler axioms, for example, there are no interacting models known in 3+1 dimensions\cite[p. 8ff]{BrunettiEtAl}.
We will here implement a reverse variant of Kadanoff's nonlinear process of \emph{blocking} spin operators to give spin operators that are associated with a larger region of space-time, followed by a rescaling process\cite{Kadanoff}.
Instead, source functions will be locally but nonlinearly transformed and separated into possibly overlapping \emph{fragments} that will be used to construct new operators.

Section \ref{Nonlinearity} begins with a very brief summary of a generating functional approach to Lagrangian quantum field theory, followed by an argument in favor of nonlinearity that explicitly includes a specification of the blocking and scaling that is required for regularization and renormalization (\ref{RealSpaceRenormalization} gives an auxiliary real-space argument.)
Section \ref{Frag} proves a corollary to the Reeh-Schlieder theorem that allows a top-down fragmentation approach to be taken seriously and begins a consideration of a moderately general ansatz that is continued in Section \ref{Cutoffs} and discussed in more general terms in Section \ref{Discussion}. 
Section \ref{ConvexHull} notes that the introduction of nonlinearity suggests a convex hull modification of microcausality.

We will work in a manifestly Poincar\'e invariant setting of the Wightman axioms on an underlying Minkowski space $\mathcal{M}$, in which free scalar quantum fields are operator-valued distributions, $\hat\phi(x)$, with free quantum field operators constructed by linearly \emph{smearing} $\hat\phi(x)$ with \emph{sources}, $\hat\phi_{\source}\,{=}\int_{\mathcal{M}}\hat\phi(x)\source(x)\rmd^4x$, for which see \ref{WightmanAxioms} and \cite[Ch. II]{Haag}.
In mathematical physics literature such as \cite{Haag}, what are here called \emph{sources} are usually called, more abstractly, \emph{test functions}.
In a previous article\cite{Morgan2019}, I have more concretely called them \emph{sampling functions} or \emph{window functions}, by analogy with a similar idea in signal analysis, or \emph{modulation functions}, insofar as $\hat\phi_{\source}|0\rangle$ is a multiplicative modulation of the vacuum vector $|0\rangle$.
The broad motivation adopted here is that the empirical evidence for quantum field theory comes to us as \emph{signals} out of many experimental apparatuses, so that it is helpful to think of quantum field theory as an idealized signal analysis formalism in the presence of different kinds of noise and using noncommutativity to model intervention contexts systematically as a path towards causal modeling\cite{Pearl}; a particle property interpretation of events that are identified on signal lines by hardware and software is countenanced only when the signal analysis of experimental data in terms of particle properties is sufficiently unambiguous.

The central idea of this article is that an analysis of renormalization gives us good reason to take quantum field operators $\hat\phi_{\source}$ to be \emph{nonlinear} functionals of the source functions (which for window functions or modulation functions in signal analysis would be a natural starting point, with linearity only adopted as an approximation), so that in general we do not work with operator-valued distributions.
A well-controlled space of source functions allows us to use powers of source functions in a well-defined way straightforwardly, whereas using powers of operator-valued distributions requires significantly more elaborate mathematics.
Section \ref{AxiomaticSection} presents an axiomatic perspective that is grounded in that nonlinearity.

\section{An argument in favor of nonlinearity}\label{Nonlinearity}
A textbook use of a source field constructs $\mathcal{Z}[\source]$ as a generating functional\cite[\S 6-1-1]{IZ} for time-ordered Vacuum Expectation Values (VEVs) for interacting quantum field operators $\hat\xi_{\source}$,
$$\mathcal{Z}[\source]=\VEV{\TO{\rme^{\rmi\hat\xi_{\source}}}}=\frac{\VEV{\TO{\rme^{\rmi S[\hat\phi]+\rmi\hat\phi_{\source}}}}}{\VEV{\TO{\rme^{\rmi S[\hat\phi]}}}},$$
from which an $n$-linear time-ordered VEV can be obtained by functional differentiation,
$$\left.\frac{1}{\rmi^n}\frac{\delta}{\delta\source(x_1)}\cdots\frac{\delta}{\delta\source(x_n)}\mathcal{Z}[\source]\right|_{\source=0}=Z(x_1,...,x_n)=\VEV{\TO{\hat\xi(x_1)\cdots\hat\xi(x_n)}},$$
or, by polarization, in an equivalent $n$-linear source form,
$$\left.\frac{1}{n!\rmi^n}\frac{\partial}{\partial\alpha_1}\cdots\frac{\partial}{\partial\alpha_n}\mathcal{Z}\left[{\textstyle\sum}_i\alpha_i\source_i\right]\right|_{\textstyle\vec\alpha{=}\vec 0}
                                    \hspace{0.3em}=Z[\source_1,...,\source_n]=\VEV{\TO{\hat\xi_{\source_1}\cdots\hat\xi_{\source_n}}}.\hspace{5.5em}$$
We write $\hat Z_{\source}\,{=}\,\TO{\rme^{\rmi\hat\xi_{\source}}}$ as a time-ordered generating operator, for which $\mathcal{Z}[\source]\,{=}\,\VEV{\hat Z_{\source}}$.
For the interacting quantum field operator $\hat Z_{\source}$, $\mathcal{Z}[\source]\,{=}\,\VEV{\hat Z_{\source}}$ is enough by itself to construct the S-matrix, however $\mathcal{Z}[\source]$ is not enough to fix the noncommutative algebraic structure of the {\StarAlgebra} that is generated by multiple copies of the $\hat Z_{\source}$ and the vacuum state, $\VEV{\hat Z_{\source_1^{\,}}\cdots\hat Z_{\source_n^{\,}}}$, which should be expected to be significant when not working only in the asymptotic S-matrix regime.

The notation above omits the regularization and renormalization scheme that has to be used in any attempt to construct a Poincar\'e invariant interacting quantum field theory.
We will adopt a generalization of a Kadanoff-type approach of layers of blocking and scaling, so we explicitly include a specification of all the interaction parameters, blocking, and scaling that will be used, $\boldsymbol{\lambda}$, $\mathbf{B}$, and $\mathbf{S}$,
{\large$$\mathcal{Z}[\source]=\VEV{\hat Z_{\source}}=\frac{\displaystyle\VEV{\TO{\rme^{\rmi S[\hat\phi,\boldsymbol{\lambda},\mathbf{B},\mathbf{S}]+\rmi\hat\phi_{\source}}}}}
                                                                                              {\displaystyle\VEV{\TO{\rme^{\rmi S[\hat\phi,\boldsymbol{\lambda},\mathbf{B},\mathbf{S}]}}}},$$}%
which we take not to be a Poincar\'e invariant construction in general, but we will adopt the usual hope and intention that by taking an appropriate limit,
$\boldsymbol{\lambda}{\rightarrow}\boldsymbol{\lambda}_c$, $\mathbf{B}{\rightarrow}\mathbf{B}_c$, $\mathbf{S}{\rightarrow}\mathbf{S}_c$, we will obtain a Poincar\'e invariant construction.
The values used for $\boldsymbol{\lambda}$, $\mathbf{B}$, and $\mathbf{S}$ typically depend on what energies are probed by the measurements being performed, which in principle is determined by the source function $\source$, so more completely we would write $\boldsymbol{\lambda}[\source]$, $\mathbf{B}[\source]$, and $\mathbf{S}[\source]$.
Hollowood\cite[\S 1.2]{Hollowood} presents renormalization group flow in a way that is well-suited to our purposes here.
He requires, for every physical measurement $F$ in a given experiment, with interaction parameters $\boldsymbol{\lambda}(\mu)$ and characteristic length scale $\ell$, that the theoretical model must be invariant under changes of the cutoff momentum scale $\mu\rightarrow\mu'$,
$$F(\boldsymbol{\lambda}(\mu);\ell)_\mu = F(\boldsymbol{\lambda}(\mu');\ell)_{\mu'},\qquad\qquad\quad\mu,\mu'>\ell^{-1}.\quad$$
The characteristic length $\ell$ is a physical parameter that is weakly dependent on, for example, whether an experiment probes electronic, atomic, nuclear, or gravitational phenomena.
In a less idealized theoretical model the physical measurement result $F$ would be determined by \emph{much} more geometrical detail about the experiment than is given by $\ell$, which here will be taken to be provided by some set of source functions $\source_1, ..., \source_n$.
The characteristic length $\ell$ would also be, although weakly, some functional of the same source functions, $\ell(\source_1, ..., \source_n)$, giving
$$F(\boldsymbol{\lambda}(\mu);\source_1, ..., \source_n)_\mu = F(\boldsymbol{\lambda}(\mu');\source_1, ..., \source_n)_{\mu'},\qquad\mu,\mu'>\ell^{-1}(\source_1, ..., \source_n).$$
In this construction, $\mu$ and $\mu'$ declare the characteristic momentum scales of two theoretical models, in contrast to the \emph{physical} momentum scale $\ell^{-1}(\source_1, ..., \source_n)$.
$\mu$ and $\mu'$ are only weakly constrained by the inequality with $\ell^{-1}(\source_1, ..., \source_n)$, however a more systematic approach would be to give the choice of $\mu$ and $\mu'$ as different functionals of the source functions, $\mu(\source_1, ..., \source_n)$ and $\mu'(\source_1, ..., \source_n)$, with both still satisfying 
$\mu(\source_1, ..., \source_n),\mu'(\source_1, ..., \source_n)>\ell^{-1}(\source_1, ..., \source_n)$, so that the renormalization group flow can be rewritten as
$$F(\boldsymbol{\lambda}(\mu(\source_1, ..., \source_n));\source_1, ..., \source_n)_{\mu(\source_1, ..., \source_n)}
     = F(\boldsymbol{\lambda}(\mu'(\source_1, ..., \source_n));\source_1, ..., \source_n)_{\mu'(\source_1, ..., \source_n)},$$
or, more succinctly, removing $\mu(\source_1, ..., \source_n)$ and $\mu'(\source_1, ..., \source_n)$ as intermediaries,
$$F(\boldsymbol{\lambda}(\source_1, ..., \source_n);\source_1, ..., \source_n) = F'(\boldsymbol{\lambda}'(\source_1, ..., \source_n);\source_1, ..., \source_n).$$
Removing $\boldsymbol{\lambda}(\source_1, ..., \source_n)$ and $\boldsymbol{\lambda}'(\source_1, ..., \source_n)$ as the final intermediaries, we obtain
$$F(\source_1, ..., \source_n) = F'(\source_1, ..., \source_n),$$
which is trivial except that $F(\source_1, ..., \source_n)$ is now definitely a nonlinear functional of the source functions.

We separate interaction parameters and blocking from scaling because blocking is essentially the complicated but relatively straightforward construction of new observables as arbitrary nonlinear functions of other observables, whereas we can take the effect of a scaling operation to be equivalent to introducing a free quantum field that has a different mass, because free scalar quantum fields are indexed by just mass as an inverse length scale.
We could define ``scaling'' in a way that includes aspects of interaction parameter scaling and of blocking, but it is helpful here to define scaling as purely about introducing free quantum fields that have a different mass.
Without loss of generality, we can replace any collection of such scaling operations that we use for a particular regularization and renormalization scheme by a possibly infinite collection of independent free quantum fields $\{\hat\phi^{(i)}\}$ of different masses, for which
$\VEV{\hat\phi_\sourcef^{(i)*}\hat\phi_\sourceg^{(j)}}\,{=}\,\delta_{i,j}(\sourcef,\sourceg)^{(i)}$, so that
{\large$$\mathcal{Z}[\source]=\VEV{\hat Z_{\source}}=\frac{\displaystyle\VEV{\TO{\rme^{\rmi S[\{\hat\phi^{(i)}\},\boldsymbol{\lambda}[\source],\mathbf{B}[\source]]+\rmi\hat\phi^{(0)}_{\source}}}}}
                                                                                              {\displaystyle\VEV{\TO{\rme^{\rmi S[\{\hat\phi^{(i)}\},\boldsymbol{\lambda}[\source],\mathbf{B}[\source]]}}}},$$}
where we have taken $\hat\phi^{(0)}_{\source}\,{=}\,\hat\phi_{\source}$.

A real-space renormalization argument, given in \ref{RealSpaceRenormalization}, is more ad-hoc but also supports the idea that the nonlinearity suggested here should be thought a significant aspect of quantum field theory.

\section{Fragmentation of the source}\label{Frag}
The {\StarAlgebra} generated by the collection $\{\hat\phi^{(i)}\}$ and the vacuum state together allow the construction of a Wightman field,
so the Reeh-Schlieder theorem\cite[\S II.5.3]{Haag}\cite[Thm. 4-2]{SW}\cite[\S II]{WittenSW} asserts that
\begin{quote}\emph{the set of Hilbert space vectors that can be constructed using the operators associated with a bounded region in space-time is dense in the set of states that can be constructed using the whole algebra.}
\end{quote}
In particular, if $\hat Z_{\source}|0\rangle$ is of finite norm, $\VEV{\hat Z_{\source}^\dagger\hat Z_{\source}^{\,}}<\infty$, which we can take to be satisfied before a limit $\boldsymbol{\lambda}[\source]{\rightarrow}\boldsymbol{\lambda}_c$, $\mathbf{B}[\source]{\rightarrow}\mathbf{B}_c$ is taken, then we can approximate $\hat Z_{\source}|0\rangle$ arbitrarily closely in the Hilbert space norm for any $\source$ by using an operator $\hat\zeta_{\source}$ that is constructed using only operators that are associated with the support of that $\source$,
$$\hat\zeta_{\source}|0\rangle \equiv \hat Z_{\source}|0\rangle,\qquad
         \left\|(\hat Z_{\source}-\hat\zeta_{\source})|0\rangle\right\|^2=\VEV{(\hat Z_{\source}-\hat\zeta_{\source})^\dagger(\hat Z_{\source}-\hat\zeta_{\source})}<\epsilon^2,$$
so that for the vacuum component of $(\hat Z_{\source}-\hat\zeta_{\source})|0\rangle$ we have $\left|\VEV{(\hat Z_{\source}-\hat\zeta_{\source})}\right|\le\left\|(\hat Z_{\source}-\hat\zeta_{\source})|0\rangle\right\|<\epsilon$ for any $\epsilon>0$.
For the $n$-linear source form of the VEVs, we can construct an operator $\hat\zeta_{{\textstyle\sum}_i\alpha_i\source_i}$ for which
$$\left.\frac{1}{n!\rmi^n}\frac{\partial}{\partial\alpha_1}\cdots\frac{\partial}{\partial\alpha_n}
                      \VEV{(\hat Z_{{\textstyle\sum}_i\alpha_i\source_i}-\hat\zeta_{{\textstyle\sum}_i\alpha_i\source_i}})\right|_{\textstyle\vec\alpha{=}\vec 0} \equiv 0.$$
We can give this as an explicit limit for $n=1$, taking $\epsilon$ to be a function of $\alpha_1$ that approaches zero faster than $\alpha_1$, as
$$\lim_{\alpha_1\rightarrow 0}
                      \frac{\left|\VEV{((\hat Z_{\alpha_1\source_1}-\hat\zeta_{\alpha_1\source_1})-(\hat Z_0-\hat\zeta_0))}\right|}{\alpha_1}
                                     \le\lim_{\alpha_1\rightarrow 0}\frac{2\epsilon(\alpha_1)}{\alpha_1} = 0,$$
and similarly but at greater length for $n>1$, so that for the purposes of the $n$-linear VEVs, and hence for the purposes of the S-matrix, we can approximate $\mathcal{Z}[\source]$ by $\VEV{\hat\zeta_{\source}}$.
We have proved
\begin{corollary}[to Reeh-Schlieder]\label{CRS}
  If $\hat Z_\source|0\rangle$ is of finite norm, the derivatives at $\source=0$ of the time-ordered generating functional $\mathcal{Z}[\source]=\VEV{\hat Z_\source}$ can be approximated arbitrarily closely by derivatives at $\source=0$ of $\VEV{\hat\zeta_\source}$, using operators $\hat\zeta_\source$ that are constructed using only functions that have support in $\Supp{\source}$.
\end{corollary}
Less than perfectly matching the S-matrix that is generated by $\mathcal{Z}[\source]$ can be good enough as an empirically adequate model, because recorded experimental data is always finite in number and precision.
Our use of $\hat\zeta_{\source}$ is also not restricted to only the asymptotic case of the S-matrix: we can also consider VEVs such as $\VEV{\hat\zeta_{\source_1^{\,}}\cdots\hat\zeta_{\source_n^{\,}}}$.

The requirement that $\epsilon(\alpha_1)$ approaches zero faster than $\alpha_1$ expresses that it may be more difficult to ensure that the derivatives of $\VEV{\hat\zeta_{\source}}$ are close to the derivatives of $\VEV{\hat Z_{\source}}$ than to ensure that $\hat\zeta_{\source}|0\rangle$ is close to $\hat Z_{\source}|0\rangle$.
That such closeness can be achieved using local operators is good to know, but better to know is whether intuitively useful and empirically successful models can be constructed using $\hat\zeta_{\source}$.
That empirical success is sufficient is important because closeness in the Hilbert space norm, $\left\|(\hat Z_{\source}-\hat\zeta_{\source})|0\rangle\right\|<\epsilon$, is not the same as closeness in the operator norm, $\left\|\hat Z_{\source}-\hat\zeta_{\source}\right\|<\epsilon$, the possibility of which is \emph{not} assured by the Reeh-Schlieder theorem.
Such closeness in the operator norm, however, is not probed experimentally by measurement of the S-matrix.

Different or more general constructions may be found to be more useful in future, but as an investigation of some elementary properties of $\hat\zeta_{\source}$ we will here adopt an ansatz that somewhat parallels the construction of $\hat Z_{\source}$ as 
$$\hat\zeta_{\source} = \TO{\rme^{\rmi\hat\phi^{(0)}_{\source} + \rmi\,\hat{\mathbf{F}}_{\source}[\{\hat\phi^{(i)}\}] }},\mbox{\quad where\quad}\hat{\mathbf{F}}_{\source}[\{\hat\phi^{(i)}\}]=\sum_n\prod_{r=1}^{R_n}\hat\phi^{(i_{(n,r)})}_{F_{(n,r)}[\source]}$$
is an operator constructed in the Hilbert space generated by $\{\hat\phi^{(i)}\}$ as some number of terms, each of which is a product of some number $R_n$ of operator-valued factors that are nonlinearly dependent on $\source$.
Instead of deforming the dynamics between different local measurements, we deform the local measurements with the intention that the deformed measurements will have the same effect as the deformed dynamics.

The support of each of the functions $F_{(n,r)}[\source]$ is required to be a subset of the support of the source function $\source$,
$\Supp{F_{(n,r)}[\source]}\subseteq\Supp{\source}$, which ensures that $[\hat\zeta_\sourcef,\hat\zeta_\sourceg]\,{=}\,0$ whenever the sources $\sourcef$ and $\sourceg$ have space-like separated supports.
We can use any reasonable \emph{non}local functionals of $\source$ freely, provided we then use them as part of a product with $\source$, insofar as $\Supp{\source{\cdot}\mbox{\sf anyF}[\source]}\subseteq\Supp{\source}$.

\begin{figure}
\includegraphics[width=1\textwidth]{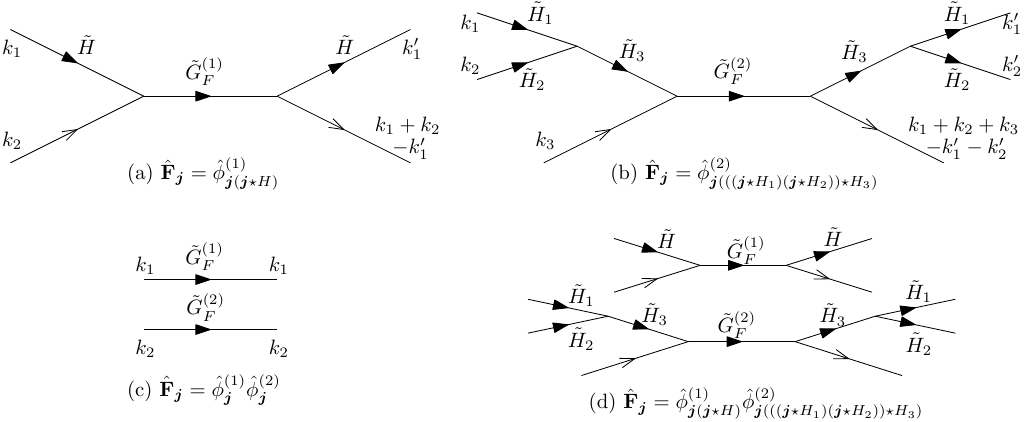}
\caption{Feynman-like diagrams for source-with-convolved-source interactions.
Each arrow is associated with a wavenumber $k$: an open arrow indicates a trivial propagator; a closed arrow indicates that the adjacent nontrivial propagator is applied to $k$, giving a factor such as $\tilde H(k)$.
Incoming wavenumbers at each vertex sum to zero.\label{FeynmanLike}}
\end{figure}

We have not so far assumed or required that $\mathcal{Z}[\source]$  is Poincar\'e invariant and well-defined, however it is straightforward to construct Poincar\'e invariant examples for $\hat\zeta_\source$.
Whether a well-defined Poincar\'e invariant construction is possible or not in a Lagrangian approach, we can still investigate the properties
of $\mathcal{Z}^{\mbox{\scriptsize(nl)}}[\source]\,{=}\,\VEV{\TO{\rme^{\rmi\hat\phi_{\source} + \rmi\,\hat{\mathbf{F}}_{\source}[\{\hat\phi^{(i)}\}] }}}$ when this alternative construction is manifestly Poincar\'e invariant.
For perhaps the most elementary construction, $\hat{\mathbf{F}}_{\source}\,{=}\,\hat\phi^{(1)}_{\source(\source{\star}H)}$, using a product of the original source with a convolution of the source with an arbitrary real-valued Lorentz invariant function $H$, $[\source(\source{\star}H)](x)\,{=}\,\source(x)\!\int\!H(x-y)\source(y)\rmd^4y$, we obtain a contribution to the 4-linear VEV, in point and source forms respectively, from the Feynman propagator term $(\source(\source{\star}H),\source(\source{\star}H))^{(1)}_F$,
\begin{eqnarray*}
  Z^{\mbox{\scriptsize(nl)}}(x_1,x_2,x_3,x_4)&=&H(x_1-x_2)G^{(1)}_F(x_2-x_3)H(x_3-x_4)+\mbox{permutations of }[x_1,x_2,x_3,x_4],\vspace{-1ex}\\
  Z^{\mbox{\scriptsize(nl)}}[\source_1,\source_2,\source_3,\source_4]&\!=\!&\!\!\!\int\!\source_1(x_1)H(x_1-x_2)\source_2(x_2)G^{(1)}_F(x_2-x_3)\source_3(x_3)H(x_3-x_4)\source_4(x_4)\rmd^4x_1\rmd^4x_2\rmd^4x_3\rmd^4x_4
\hspace{-17.2em}\mbox{\raisebox{-3.5ex}{$+\mbox{ permutations of }[\source_1,\source_2,\source_3,\source_4],$}}
\end{eqnarray*}
which is an interaction of \emph{some} kind, but specified in an algebraic way instead of by an interaction dynamics and a regularization and renormalization scheme.
In wavenumber space, this expression can be presented in a Feynman diagram-like way as in Figure \ref{FeynmanLike}(a), where one of the arrows must correspond to  a trivial propagator to ensure microcausality is satisfied.
We can extend this particular construction to any tree-level diagram that has at least one trivial propagator, as for
$\hat{\mathbf{F}}_{\source}\,{=}\,\hat\phi^{(2)}_{\source(((\source\star H_1)(\source\star H_2))\star H_3)}$, which gives a contribution to the 6-linear VEV that is presented graphically in Figure \ref{FeynmanLike}(b).
This construction is local because the test function $\source(((\source\,{\star}\,H_1)(\source\,{\star}\,H_2))\,{\star}\,H_3)$ includes $\source$ as a factor, however the factor $((\source\,{\star}\,H_1)(\source\,{\star}\,H_2))\,{\star}\,H_3$ represents a propagation of $\source$ to a point $y$ that may be anywhere in space-time, in two different ways, giving us $[\source\,{\star}\,H_1](y)[\source\,{\star}\,H_2](y)$, then a propagation of that value at every point back to the support of $\source$.
Such constructions are not modified by the presence of any other measurements, but they represent a first step towards introducing many different resonances between sources and towards interference effects between those different resonances.

The measurement operator constructions given above work with fragments of the source  at a point, $\source(x)$, nonlinearly modulating those fragments by multiplication with other functions, but despite the nontrivial $n$-linear VEVs that result we nonetheless obtain a structure that is still Gaussian insofar as the most elementary case above (with the vacuum constructed using raising and lowering operators as usual) is generated by $\hat\varphi_{\source}=\hat\phi^{(0)}_{\source}+\hat\phi^{(1)}_{\source(\source{\star}H)}$, which satisfies the Heisenberg algebra,
$$[\hat\varphi_\sourcef,\hat\varphi_\sourceg]=(\sourcef^*,\sourceg)^{(0)}-(\sourceg^*,\sourcef)^{(0)}
       +((\sourcef(\sourcef{\star}H))^*,\sourceg(\sourceg{\star}H))^{(1)}-((\sourceg(\sourceg{\star}H))^*,\sourcef(\sourcef{\star}H))^{(1)}.$$
More in the spirit of Kadanoff's nonlinear ``blocking'' of spins, however, we can construct products of two or many more such measurement operator constructions, which will create a non-Gaussian quantum field.
We can, most simply, introduce $\hat{\mathbf{F}}_{\source}\,{=}\,\hat\phi^{(1)}_{\source}\hat\phi^{(2)}_{\source}$, as in Figure \ref{FeynmanLike}(c).
This construction ensures that the algebra generated by $\hat\zeta_\source$ is a strict subalgebra of the Wightman field {\StarAlgebra} generated by $\hat\phi^{(0)}_{\source}$, $\hat\phi^{(1)}_{\source}$, and $\hat\phi^{(2)}_{\source}$, because $\hat\phi^{(1)}_{\source}$ and $\hat\phi^{(2)}_{\source}$ always occur together, which introduces interference effects because of products such as $(\source_1,\source_2)^{(1)}(\source_3,\source_4)^{(2)}$.

As in Figure \ref{FeynmanLike}(d) (which first introduces any modification of the quantum field model when we measure 10-linear VEVs), we can introduce products of operators that use any local functional of the source function, which implicitly generalizes the concept of a ``fragment'', somewhat extremely, to be \emph{any} functional $F[\source]$ for which $\Supp{F[\source]}\subseteq\Supp{\source}$.
Note, however, that local functionals are possible that defy easy graphical presentation, such as $\source(\source{\star}\source{\star}H)$, or, using derivatives, complex conjugation, and arbitrary powers, such as
$\partial_\mu\source\,{\cdot}\,\partial^\mu(\source{\star}H)$,
$\partial_\mu\partial_\nu\source\,{\cdot}\,\partial^\mu(\source{\star}H_1)\,{\cdot}\,\partial^\nu(\source{\star}H_2)$,
$|\source|^\alpha\,{\cdot}\,(\source{\star}H)$, \emph{et cetera}, or we can introduce scaling effects such as $F[\source](x)=\source(x)\source(\mu{\cdot}(x-x_\source)+x_\source)$, where $x_\source$ is a constant weighted center of the source function and $\mu$ is a constant scaling factor.
All such nonlinear functionals, all satisfying microcausality, emphasize some source functions over others in their own way, resulting in different kinds of resonance.
If we construct a continuum of slightly different resonances, then we can expect the interference between them sometimes to present as caustics.


As constructed here, we can introduce new expressions that have an effect only for $n$-linear VEVs for large $n$, but that have no effect whatsoever on lower order VEVs.
We could also consider an ansatz, however, that more parallels the usual definition of an interacting field\cite[Eq. (6-5)]{IZ}, $\hat\xi(x)\,{=}\,\hat U^{-1}(t)\hat\phi(x)\hat U(t)$, such as
$\hat\zeta_\source\,{=}\,\TO{\rme^{\rmi\hat{\mathbf{F}}_{\source}[\{\hat\phi^{(i)}\}]}}^\dagger\!\rme^{\rmi\hat\phi_\source}
                                           \TO{\rme^{\rmi\hat{\mathbf{F}}_{\source}[\{\hat\phi^{(i)}\}]}}$, with $\hat{\mathbf{F}}_{\source}[\{\hat\phi^{(i)}\}]$ as a nonlinear local functional of the source function, but such a construction reduces by use of a Baker-Campbell-Hausdorff identity to a form similar to that considered above.
It will be apparent that the ansatz we have discussed here is effectively a generating functional for tree-level diagrams, with very few constraints, so it is somewhat close to the Feynman Tree Theorem\cite{FeynmanGravity,FeynmanTreeTheorem} and Generalized Unitarity\cite{BernHuang}, however the approach here adopts a very different philosophy of using nonlinear functionals of the source function as a way to sidestep the concerns of regularization and renormalization, with a much looser connection, through the Reeh-Schlieder theorem, to any specific Lagrangian.

\section{Cutoffs and other meromorphic functions}\label{Cutoffs}
All the constructions in the previous section use only multinomials in components of the source functions, which we can extend straightforwardly to entire functions.
If we introduce a cutoff, however, in a construction such as $\hat{\mathbf{F}}_{\source}[\{\hat\phi^{(i)}\}]\,{=}\,\hat\phi^{(1)}_{\tanh(\lambda\source)}$, $\hat\zeta_{\source}\,{=}\,\TO{\rme^{\rmi(\hat\phi^{(0)}_{\source}+\hat\phi^{(1)}_{\tanh(\lambda\source)})}}$ generates a more complicated system of $n$-point operator-valued distributions because $\source(x)$ may take values outside the domain of convergence of expansions of $\tanh(\lambda\source(x))$ at $\source(x)=0$, in which case the derivatives at zero source field that give the VEVs of the model will not give complete information about $\hat\zeta_{\source}$.
For $\tanh(\lambda\source(x))$, or for any other nontrivial meromorphic function, it will instead be necessary to use analytic continuation.

Even more significantly, if two source functions pass on different sides of a pole of $\tanh(\lambda\source(x))$, then they cannot be continuously deformed to be equal to each other without passing through that pole, in which case the norm of the Hilbert space vector $\TO{\rme^{\rmi(\hat\phi^{(0)}_{\source}+\hat\phi^{(1)}_{\tanh(\lambda\source)})}}|0\rangle$ would be undefined, thereby introducing a discrete structure that is determined by the poles of the meromorphic functions that are used.
Cutoffs are a commonplace in experiment and in classical signal analysis and engineering and cannot be ruled out \emph{a priori} in the quantum field case.

\section{An axiomatic perspective}\label{AxiomaticSection}
The only change that is needed to the presentation of free Wightman fields in \ref{WightmanAxioms} is the removal of complex linearity, so that $\hat\xi_{\lambda\sourcef+\mu\sourceg}\,{\not=}\,\lambda\hat\xi_\sourcef+\mu\hat\xi_\sourceg$.
In other words, $\hat\xi:\mathcal{F}\rightarrow\mathcal{A};\sourcef\mapsto\hat\xi_\sourcef$ is \emph{not} an operator-valued distribution.
The measurement algebra is still generated by $\hat\xi_\sourcef$ and still satisfies microcausality, and the vacuum state is still a linear map that allows the construction of a Hilbert space $\mathcal{H}$ that supports a representation of the Poincar\'e group and for which the joint spectrum of the generators of the action of translations on $\mathcal{H}$ is a 4-vector in the (closed) forward light-cone.

No weakening of the linearity of the Wightman field is mentioned in Streater's relatively early review\cite[\S 3.4]{Streater}, but in the context of the Haag-Kastler axioms a recent approach to interactions introduces a form of nonlinearity that is of interest\cite{BuchholzFredenhagen}.
The approach here is much more constructive, takes renormalization as a starting point, and ---for better or worse--- preserves more contact with the textbook approach to quantum field theory, however we can adapt a construction in \cite{BuchholzFredenhagen} to suggest an intermediate axiom, instead of linearity or completely abandoning linearity, that
$$\hat\Phi_{\source_1+\source_2+\source_3}=\hat\Phi_{\source_1+\source_3}+\hat\Phi_{\source_2+\source_3}-\hat\Phi_{\source_3}\quad\mbox{whenever $\source_1$ and $\source_2$ have space-like separated supports.}$$
$\source_1(x)\source_2(x)=0\,\forall x$, if the two functions have space-like separated supports, so it is easily verified that this axiom is satisfied for $\hat\Phi_{\source}=\hat\phi_{\source^{\hspace{0.06em}n}}$, for any $n$ and for any sum of similar expressions, because
$$(\source_1+\source_2+\source_3)^n-((\source_1+\source_3)^n+(\source_2+\source_3)^n-\source_3^{\hspace{0.06em}n})$$ always includes a factor $\source_1\source_2$, so the axiom above is satisfied for the ansatz $\hat\Phi_{\source}=\hat\phi^{(0)}_{\source}+\sum_n\hat\phi^{(n)}_{\!\source^{\hspace{0.06em}n+1}}$, for example; but it is only satisfied for $\hat\Phi_{\source}=\hat\phi_{\!\source^{\hspace{0.06em}n}{\cdot}(\source{\star}H)}$ if $H$ is a weighted sum of advanced and retarded propagators, because all terms in the difference include one of the factors $\source_1\source_2$, $\source_1(\source_2{\star}H)$, or $(\source_1{\star}H)\source_2$.
The axiom above is not satisfied, however, by $\hat\Phi_{\source}=\hat\phi_{\source^{\hspace{0.06em}m}}\hat\phi_{\source^{\hspace{0.06em}n}}$, for example, which at this point does not seem obviously ruled out by any empirical principle.
For this latter construction, therefore, more elaborate axiomatic constraints would be required.

The nonlinearity of $\hat\xi_\sourcef$ has the drastic consequence that it allows $n$-linear quantum field operators to be generated by functional differentiation, giving $n$-point operator-valued distributions $\hat\xi(x_1, ...,x_n)$, as well as more complex behavior if $\hat\xi_\sourcef$ is a meromorphic functional of $\sourcef$.
Such constructions are in this formalism clearly a consequence of nonlinearity, and of dispersion if derivatives are introduced nontrivially, which may be understood as bound states if the interaction is attractive, but may make more general contributions to global behavior.
If we did not have the strong suggestion from the analysis of renormalization in Section \ref{Nonlinearity} that nonlinearity is hidden but is in fact present in the formalism as we currently have it, the consequence of $n$-point operator-valued distributions $\hat\xi(x_1, ...,x_n)$ would seem a remarkable innovation.

The Reeh-Schlieder theorem can seem quite counter-intuitively nonlocal, however it is a natural consequence of the boundary conditions implicit in the way the Poincar\'e group is used in the Wightman axioms.
As an intuition that perhaps better accommodates that nonlocality, using a local operator to modify a state as an approximate model of the world modulates not only measurement results in the future but also measurement results in the past and present to be {\sl consistent} with that modulated future, which then modifies measurement results even outside the support and causal future of the local operator that was used.
There is, loosely, something of a back-and-forth of \emph{consistency}, however whether we label that consistency ``causality'' is a delicate question (which apart from this paragraph will be taken to be outside the scope of this article).
The construction used in Witten's proof of the Reeh-Schlieder theorem\cite{WittenSW} can be read as giving a precise meaning to this consistency in terms of complex analyticity.
Despite this back-and-forth of consistency, quantum field theory nonetheless only includes modulations of the statistics of measurement results that satisfy microcausality, which ensures that messages cannot be sent.

Even when thinking in terms of classical probability our intuition should take any question we ask to be nonlocal, insofar as most of what happened to us before we ask the question is available to someone who is space-like separated from us.
If someone knows a lot about my past until a minute before I ask a question, they can make an informed guess about what the question might be.
They don't know exactly, but quantum field theory and similar classical theories are not about certainties, they are about probabilities in Minkowski space, so of course they encode, in a very abstract way, that we can make better than random guesses about what happens at space-like separation because we know something about the past.

\section{Convex Hull Microcausality}\label{ConvexHull}
Two functions satisfy \emph{Convex Hull Microcausality} if the convex hulls of the supports of the two functions satisfy microcausality.
Convex hull microcausality is a weaker requirement than microcausality when the support of at least one of the functions is not convex, in which case we can say that one of the functions may ``surround'' the other, as for a ring around a cylinder, or that the functions may surround each other, as for two interlinked rings.
The concept of convex hull microcausality has not to my knowledge been previously introduced in the quantum field theory literature, so experimental tests that could distinguish between microcausality and convex hull microcausality have not been discussed.
At first sight, it seems that a definitive test might be rather difficult.

To show how convex hull microcausality emerges as a rather natural definition that is only slightly weaker than microcausality when we introduce nonlinearity, we first recall that we can construct a Gaussian Wightman field as a sum of raising and lowering operators, $\hat\xi_f=a_{f^*}+a_f^\dagger$, for which
$$[a_\sourcef,a_\sourceg^\dagger]=(\sourcef,\sourceg)=\int \tilde\sourcef^*\!(k)\tilde G(k)\tilde\sourceg(k)\kInt{k},\qquad\tilde G(k)=2\pi\delta(k{\cdot}k\,{-}\,m^2)\theta(k_0),$$
then for the measurement operator commutator we have
\begin{eqnarray*}
[\hat\xi_\sourcef,\hat\xi_\sourceg]&=&(\sourcef^*,\sourceg)-(\sourceg^*,\sourcef)
              =\int\left(\tilde\sourcef(-k)\tilde\sourceg(k)-\tilde\sourceg(-k)\tilde\sourcef(k)\right)\tilde G(k)\kInt{k}\cr
      &=&\int\left(\rme^{\rmi k\cdot(x-y)}-\rme^{-\rmi k\cdot(x-y)}\right)\tilde G(k)\kInt{k}\cdot\sourcef(x)\sourceg(y)\rmd^4x\rmd^4y,
\end{eqnarray*}
which is zero whenever every point in $\sourcef$ is space-like separated from every point in $\sourceg$ because in that case $(x-y)$ is always space-like so that the $k$ integrand is antisymmetric.
If we require only convex hull microcausality, however, we can introduce a Gaussian nonlinear Wightman field as a sum of raising and lowering operators as above, for which, for example,
$$[a_\sourcef,a_\sourceg^\dagger]=\int\widetilde{F[\sourcef]}^*\!(k)\tilde G(k)\widetilde{F[\sourceg]}(k)\kInt{k},\quad\mbox{with}\quad
       \widetilde{F[\sourcef]}(k)=\tilde\sourcef(k)^n,$$
corresponding to an $n$-fold real-space convolution.
For the measurement operator commutator we now have
\begin{eqnarray*}
[\hat\xi_\sourcef,\hat\xi_\sourceg]&=&(\sourcef^*,\sourceg)-(\sourceg^*,\sourcef)
     =\int\left(\tilde\sourcef(-k)^n\tilde\sourceg(k)^n-\tilde\sourceg(-k)^n\tilde\sourcef(k)^n\right)\tilde G(k)\kInt{k}\cr
      &=&\int 2\rmi\sin{\left(k\cdot\left(\sum_{j=1}^n x_j-\sum_{j=1}^n y_j\right)\right)}\tilde G(k)\kInt{k}\cdot\prod_{j=1}^n\sourcef(x_j)\sourceg(y_j)\rmd^4x_j\rmd^4y_j,
\end{eqnarray*}
which is zero if $\sourcef$ and $\sourceg$ satisfy convex hull microcausality because in that case $\frac{1}{n}\sum_{j=1}^n x_j$ is in the convex hull of $\sourcef$ and $\frac{1}{n}\sum_{j=1}^n y_j$ is in the convex hull of $\sourceg$, so that $\sum_{j=1}^n x_j-\sum_{j=1}^n y_j$ is space-like.
More generally, we can scale each component of the real-space convolution, $\widetilde{F[\sourcef]}(k)=\prod_{j=1}^n\tilde\sourcef(\alpha_jk)$, $\alpha_j>0$, for which we obtain
$$[\hat\xi_\sourcef,\hat\xi_\sourceg]=\int 2\rmi\sin{\left(k\cdot\left(\sum_{j=1}^n \alpha_jx_j-\sum_{j=1}^n \alpha_jy_j\right)\right)}\tilde G(k)\kInt{k}\cdot
                                                                   \prod_{j=1}^n\sourcef(x_j)\sourceg(y_j)\rmd^4x_j\rmd^4y_j,
$$
which is again zero if $\sourcef$ and $\sourceg$ satisfy convex hull microcausality because in that case $\sum_{j=1}^n \alpha_jx_j/\sum_{j=1}^n \alpha_j$ is in the convex hull of $\sourcef$ and $\sum_{j=1}^n \alpha_jy_j/\sum_{j=1}^n \alpha_j$ is in the convex hull of $\sourceg$.
Even more generally, we can introduce $\widetilde{F[\sourcef]}(k)=\prod_{j=1}^n\widetilde{F_j[\sourcef]}(\alpha_jk)$, where $\Supp{F_j[\sourcef]}\subseteq\Supp{\sourcef}$ for each $F_j[\sourcef]$, with the same result, and we can construct $\widetilde{F[\sourcef]}(k)$ as a sum of such terms provided $\sum_{j=1}^n \alpha_j$ has the same value for each term.

\section{Discussion}\label{Discussion}
Wilson\&Kogut gives a telling summary of the physics of quantum field theory\cite[p. 79]{WilsonKogut}, \guillemotleft the behavior of the system is determined primarily by the fact that there is cooperative behavior, plus the nature of the degrees of freedom themselves. The interaction Hamiltonian plays only a secondary role.\guillemotright\ 
A slightly different way to refer to ``cooperative behavior'', motivated by signal analysis, is to think of each nonlinear transform $F_{\!(n,r)}[\source]$ as an \emph{antenna functional} $F_{\!(n,r)}$ applied to the source function to give an \emph{antenna function} (which seems an idiosyncratic enough term that we have called it an abstract fragment elsewhere here).
We can consider a source function from the perspective of each of many different antenna functionals, then we can consider to what extent the antenna functions we obtain resonate with antenna functions elsewhere, according to the corresponding dynamics, given by Wick contraction of $\hat\phi^{(i)}_{F[\source]}$ with similar terms.
In contrast to summing Feynman path weights over all possible paths, \textbf{Corollary \ref{CRS}} allows us to construct a systematic weighted sum of products of all such resonances.
Much more generally than Kadanoff's ``blocking'', we can construct multiple overlapping antennae, effectively operating at different wavenumbers and at different ranges of mass.
For a radio antenna, for example, the whole assembly transmits in the radio spectrum in relatively focused directions, but fragments of the antenna also emit thermal radiation in the infrared spectrum, more-or-less omnidirectionally, and there are electromagnetic emissions in all frequency ranges.
Figure \ref{FeynmanLike}(d), for example, can be thought of as a model in which any source function generates two quite different kinds of transmission and response.

\begin{figure}[t]
\includegraphics[width=1\textwidth]{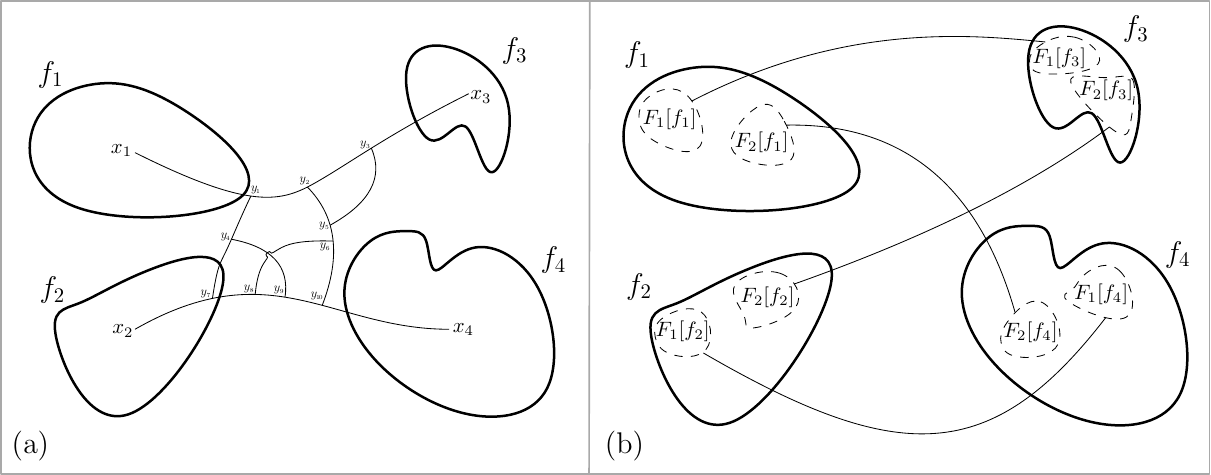}
\caption{For each of $N$ regions where we perform measurements, we have a source function $f_n$ that defines how fine-grained each measurement is and other detailed properties within that region.
Source functions are often taken to be plane waves, focused at a single wave-number, however they should in general be taken to be less singular than that.
We can think of a source function as telling us precisely how close a measurement is to an ideal but ill-defined measurement at a single point or at a single wave-number.\\[1ex]
\hspace*{\fill}
\begin{minipage}{0.4\textwidth}(a) Feynman: sum amplitudes for many Feynman diagrams, with weights determined by an algebraic structure.  Each diagram sums for every $y_i$ being everywhere in space-time and for every $x_n$ being everywhere in the support of $f_n$.
Such integrals are in general not well-defined unless we decide on a regularization or renormalization scheme that depends on the experiment and hence on details of the source functions that describe it.\end{minipage}\hspace*{\fill}
\begin{minipage}{0.49\textwidth}(b) Fragmentation: sum amplitudes for many fragment pairings, with weights determined by an algebraic structure.  Each fragment pairing gives a product of ``resonance overlaps'' $(F_j[f_m], F_j[f_n])_j$, using fragment functionals $F_j$ to construct a collection of fragments $F_j[f_n]$ for each source function.
This construction can be thought of as addressing a localized inverse problem for the interacting Feynman dynamics: the fragment functionals and resonance overlaps are required to give an equivalent overall sum, but using only a collection of free field dynamics and interference between them. \end{minipage}
\hspace{0.01\textwidth}\label{FeynmanComparison}}\vspace{-1.5ex}
\end{figure}
The construction here is a principled replacement for Lagrangian approaches to quantum field models, because of \textbf{Corollary \ref{CRS}}, which asserts that we can solve a localized inverse problem for the interacting dynamics, as in Figure \ref{FeynmanComparison}.
We can hope this will be effective as an engineering tool, but it is arguably not constrained or explanatory enough, so it is presented here more as a stimulus for other constructions than as an end in itself.
The construction is in any case different enough from Lagrangian approaches that it will take time for us to come to an understanding of its detailed consequences.
It is also clear that the wealth of possible such models and their relationship to experiment is more alike to the wealth of effective field models\cite{Georgi} than to the very restricted class of renormalizable quantum field models, which are often regarded as more fundamental because there is not such a wealth.
The approach here for example shares three characteristics with Effective Field Theories and with Schwinger's source theory that are identified by Cao\cite[p. 122]{Cao},\vspace{-1ex}
\newcounter{CaoList}
\begin{list}{\roman{CaoList}.}{\usecounter{CaoList}}{\setlength{\itemsep}{-0.5ex}
\item The denial of fundamental theory;
\item the flexibility of incorporating new particles and new interactions into existing schemes; and
\item the possibility of considering nonrenormalizable interactions.}
\end{list}\vspace{-0.5ex}
For a recent survey of Effective Field Theories in the philosophical literature, see \cite{RivatGrunbaum}, which agrees with Cao in its conclusion that \guillemotleft Effective field theories are not fundamental by intent and by design.\guillemotright\
The mathematics pursued in this article has to be taken to be mostly \emph{ad-hoc}\cite[\S 5]{Fraser} because it chooses to encode within the supports of the source functions we use what we might otherwise say ``really happens'' \emph{between} the supports of those source functions.
The mathematical desirability of a construction being well-defined does not have to be taken to be philosophically significant.

As part of a discussion of \guillemotleft Emergence as Dependence Plus Novelty and Autonomy\guillemotright, \cite[\S 2.4]{Crowther} suggests that
\begin{quote}\guillemotleft Novelty is taken as robust behaviour exhibited by the macro-system (appropriately
described by the emergent theory) but not present in the micro-system (described by
the micro theory).\guillemotright\end{quote}
A source function approach allows us to rewrite this much more explicitly as ``Novelty can be taken as robust behaviour measured at the macro-system level (appropriately
described by large scale source functions) but not present in measurements at the micro-system level (described by small-scale source functions)'', insofar as everything we know comes from measurements that we have to place relative to some coordinate system, which in a field theoretic approach uses source functions.
In a fragmentation approach, arbitrarily many or a continuum of types of fragments are constructed and their resonances brought together, so that when we say ``more is different'', that bringing together of the effects of many fragments in general will be \emph{much} more than a simple sum of one type of part.
The idea of a ``tower of emergent theories'' (see \cite[\S 2.3, \S 3.5]{Crowther}, for example), can be accommodated in a unified formalism of measurements indexed by source functions, in which all scales can be represented as needed.

There are many books and classic articles on Effective Field Theories from which many different lessons can be extracted; here we will close by contrast with one aspect of a quote from Georgi\cite[p. 214]{Georgi},
\begin{quote}\guillemotleft we can use an effective field theory to describe physics at a given energy scale, $E$, to a given accuracy, $\epsilon$, in terms of a quantum field theory with a finite set of parameters. We can formulate the effective field theory without any reference to what goes on at arbitrarily small distances.\newline
The ``finite set of parameters'' part sounds like old-fashioned renormalizability. However, the dependence on the energy scale, $E$, and the accuracy, $\epsilon$, is the new feature of effective field theory. It arises because we cannot possibly know, in principle, what is going on at arbitrarily high energies.\guillemotright
\end{quote}
A free quantum field theory and the subalgebras of free quantum field theories constructed here don't \emph{know} what is going on at arbitrarily small scales, but they do present explicit well-defined models of what {\sl would} be going on at \emph{all} scales {\sl if} the theory were completely correct (it won't be!), even for measurement statistics at such ludicrously small scales as $10^{-1000}$ meters, which can be checked and revised as needed when it is found to conflict, as it eventually will, with experiment.

As always, there are some other traditions in the literature that are somewhat comparable to the construction here of many functionals of a source function.
Two I am particularly aware of are:
\begin{itemize}
\item \emph{dressed particles}\cite{Ochiai,Ohtsu} and \emph{Infraparticles}\cite{ChenI,ChenII} surround a bare particle by a systematically generated cloud of particles.
The construction here gives an explicit functional form for how a regularization and renormalization scheme modifies the dynamics that is fixed by the interaction Lagrangian.
\item \emph{Poincar\'e invariant quantum theory}\cite{HamiltonianDynamicsA,HamiltonianDynamicsB} introduces multi-point objects as models for bound states.
The nonlinear construction here can be thought of as a generating functional for such objects, however we have insisted on microcausality here and multi-point objects are not sufficient if we introduce the meromorphic functionals of Section \ref{Cutoffs}.
\end{itemize}

The idea of an antenna gives some idea of the complexity that must be conveyed by a source function.
A description of the intuition this suggests will be attempted in this speculative final paragraph.
A source function that is mostly constant across a cubic meter volume but zero elsewhere does not, for example, tell whether it describes an iron cube or ice or some other material, nor does it describe whatever lattice defects there might be, all of which will affect how well the given source function will act as a type of antenna at different wavenumbers.
If a source function is modulated, however, at wavenumbers and amplitudes that correspond to a lattice of iron atoms, then the antenna functionals for that modulation ought to generate stronger resonances than would be generated for other materials.
We should also find that sufficiently elaborate surface and internal modulations will result in what have become known as metamaterial properties; as source functions become more complex, they might be able to model biological structures such as leaves, with it being something of an understatement, however, that the computational needs for such modeling will increase extraordinarily.
One of the most important ``transmissions'' for any source function is into the past and future: when a source is extensively modulated, its antenna functionals will hopefully transmit the object much more strongly along the worldline in Minkowski space that we would expect for a solid object, as a kind of caustic, than when a source is much less modulated, creating a distinction between source functions that behave more or less classically.
A source function or set of source functions that models a chair, say, \emph{must} create something close to this kind of classical worldline in Minkowski space for the model to be empirically adequate.
I cannot say enough, however, that thinking in terms of systematically constructing elaborate antennae and a complex of free field relationships between them feels like only a step along the way to more effective conceptualizations.

\section*{Acknowledgements}
I am grateful to Palle Jorgensen for comments on an earlier draft and to Juan R. Gonz\'alez \'Alvarez for pointing out Ref. \cite{HamiltonianDynamicsA} and clarifying some of its aspects.


\appendix
\section{The Wightman Axioms, the free scalar quantum field, and time ordering}\label{WightmanAxioms}
The Wightman axioms can be found in many variations.
As a brief paraphrase of Haag's presentation\cite[\S II.1.2]{Haag},\vspace{0.5ex}\\
\centerline{\fbox{\begin{minipage}{0.9\textwidth}\begin{itemize}{\setlength{\itemsep}{-0.5ex}
\item A Hilbert space $\mathcal H$ supports a unitary representation of the Poincar\'e group and there is a unique lowest energy Poincar\'e invariant vacuum vector $|0\rangle$.
\item Quantum fields are operator-valued distributions:\hspace{0.5em}complex linear maps from a space $\mathcal{F}$ of source functions into a {\StarAlgebra} $\mathcal A$ of operators,
$\hat\xi:\mathcal{F}\rightarrow\mathcal{A};\sourcef\mapsto\hat\xi_\sourcef$, $\hat\xi_{\lambda\sourcef+\mu\sourceg}\,{=}\,\lambda\hat\xi_\sourcef+\mu\hat\xi_\sourceg$.
\item The source functions can be Lorentz scalars, vectors, \emph{et cetera}.
\item Microcausality:\hspace{0.5em}$[\hat\xi_\sourcef,\hat\xi_\sourceg]\,{=}\,0$ if $\sourcef$ and $\sourceg$ have space-like separated supports.
\item Completeness:\hspace{0.5em}the action of the quantum field operators on $|0\rangle$ generates $\mathcal H$.
}\end{itemize}\end{minipage}}}\vspace{0.5ex}
\noindent For the free scalar quantum field, which will be our only concern here, the source function $\sourcef(x)$ is a complex scalar function on Minkowski space, $\sourcef:\mathcal{M}\rightarrow\mathbb{C}$ and $\hat\xi_\sourcef^\dagger=\hat\xi^{\ }_{\sourcef^*}$.
Taking the index set for measurement operators to be source or test functions is not necessarily the only, best, or a sufficient approach to describing experiments, however it will be taken in this article to be general enough to merit detailed consideration.

The point of this article is that we can replace just the second entry above by\vspace{0.5ex}\\
\centerline{\fbox{\begin{minipage}{0.9\textwidth}\begin{itemize}
\item Quantum fields are nonlinear maps from a space $\mathcal{F}$ of source functions into a {\StarAlgebra} $\mathcal A$ of \mbox{operators,}
$\hat\xi:\mathcal{F}\rightarrow\mathcal{A};\sourcef\mapsto\hat\xi_\sourcef$, $\hat\xi_{\lambda\sourcef+\mu\sourceg}\,{\not=}\,\lambda\hat\xi_\sourcef+\mu\hat\xi_\sourceg$,
\end{itemize}\end{minipage}}}\vspace{0.5ex}
which allows the construction of a wide range of interacting quantum field theories.

For a free scalar quantum field $\hat\phi_{\!\sourcef}^{\,}$ parameterized by mass $m$, the 2-measurement VEV
$$\VEV{\hat\phi_{\!\sourcef}^\dagger\hat\phi_\sourceg^{\,}}=(\sourcef,\sourceg)=\int\tilde\sourcef^*(k)\tilde G(k)\tilde\sourceg(k)\frac{\rmd^4k}{(2\pi)^4},\qquad\mbox{where }\tilde G(k)=2\pi\delta(k{\cdot}k-m^2)\theta(k_0),$$
is a Poincar\'e invariant positive semi-definite pre-inner product on the source function space, so for the free scalar quantum field a source function $\sourcef(x)$ must have a well-enough controlled Fourier transform $\tilde\sourcef(k)$ that $(\sourcef,\sourcef)<\infty$.
The commutator $[\hat\phi_{\!\sourcef}^{\,},\hat\phi_\sourceg^{\,}]\,{=}\,(\sourcef^*,\sourceg)\,{-}\,(\sourceg^*,\sourcef)$ is nontrivial in general but to satisfy microcausality it must be zero whenever $\sourcef$ and $\sourceg$ have space-like separated supports.
We can use Wick expansion to write down VEVs for $\rme^{\rmi\hat\phi_{\!\sourcef}^{\,}}$ and for the time-ordered expressions $\TO{\rme^{\rmi\hat\phi_{\!\sourcef}^{\,}}}$ and
$\TO{\rme^{\rmi\hat\phi_{\!\sourcef}^{\,}}}{}^{{}^{\scriptstyle\dagger}}$ as
\begin{eqnarray*}
  &&\VEV{\rme^{\rmi\hat\phi_{\!\sourcef}^{\,}}}=\rme^{-(\sourcef^*,\sourcef)/2},\qquad
  \VEV{\TO{\rme^{\rmi\hat\phi_{\!\sourcef}^{\,}}}}=\rme^{-(\sourcef,\sourcef)_F/2},\qquad
     \VEV{\TO{\rme^{\rmi\hat\phi_{\!\sourcef}^{\,}}}{}^{{}^{\scriptstyle\dagger}}}=\rme^{-(\sourcef,\sourcef)_F^*/2},\\
  &&\VEV{\rme^{\rmi\hat\phi_{\!\sourcef_1}^{\,}}\cdots\rme^{\rmi\hat\phi_{\!\sourcef_n}^{\,}}}
                             =\rme^{-\sum_i(\sourcef_i^*,\sourcef_i^{\,})/2-\sum_{i<j}(\sourcef_i^*,\sourcef_j^{\,})},\\
  &&\VEV{\TO{\rme^{-\rmi\hat\phi_{\!\sourcef_1}^{\,}}}{}^{{}^{\scriptstyle\dagger}}\TO{\rme^{\rmi\hat\phi_{\!\sourcef_2}^{\,}}}^{\,}}
          =\rme^{-(\sourcef_1,\sourcef_1)_F^*/2-(\sourcef_2,\sourcef_2)_F/2-(\sourcef_1,\sourcef_2)}, ...
\end{eqnarray*}
where $(\sourcef,\sourceg)^{\,}_F\,{=}\,\VEV{\TO{\hat\phi_\sourcef\hat\phi_\sourceg}}$ is the Feynman propagator in source function form (note that the Feynman form is unbounded, bilinear, and symmetric, in contrast to the positive semi-definiteness and sesquilinearity of $(\cdot,\cdot)$\,).
We can generalize the above to give an explicit expression for the VEV of products of any number of factors
$\rme^{\rmi\hat\phi_{\!\sourcef}^{\,}}$, $\TO{\rme^{\rmi\hat\phi_{\!\sourcef}^{\,}}}$, and $\TO{\rme^{\rmi\hat\phi_{\!\sourcef}^{\,}}}{}^{{}^{\scriptstyle\dagger}}$,
which gives a state over the {\StarAlgebra} that is generated by
$\rme^{\rmi\hat\phi_{\!\sourcef}^{\,}}$, $\TO{\rme^{\rmi\hat\phi_{\!\sourcef}^{\,}}}$, and $\TO{\rme^{\rmi\hat\phi_{\!\sourcef}^{\,}}}{}^{{}^{\scriptstyle\dagger}}$,
which in turn allows the Gelfand-Naimark-Segal-construction of a Hilbert space and a vacuum sector representation of that {\StarAlgebra}.
Within that vacuum sector representation, we have the equivalence
$$\TO{\rme^{\rmi\hat\phi_{\!\sourcef}^{\,}}}\equiv\rme^{\rmi\hat\phi_{\!\sourcef}^{\,}}\rme^{-(\sourcef,\sourcef)_F/2+(\sourcef^*,\sourcef)/2},
$$
so the vacuum sector representations of this free quantum field with or without time-ordering can be taken to be in this sense equivalent.

It is taken here to be helpful to consider a signal analysis perspective, in which a source or test function may also be thought of as
(1) a \emph{window} function, insofar as for a smeared interacting operator $\hat\phi_{\!\sourcef}$ the test function $\sourcef(x)$ specifies how a measurement couples to and hence how it ``sees'' or resonates with the state that is measured; or as
(2) a \emph{modulation} function, insofar as a smeared interacting operator $\hat\phi_\sourceg^{\,}$ can be used to modulate the vacuum state, giving states that give different expectation values, such as, at the lowest level,
$$\frac{\VEV{\hat\phi_\sourceg^\dagger\TO{\hat\phi(x_1)\cdots\hat\phi(x_n)}\hat\phi_\sourceg^{\,}}}{\VEV{\hat\phi_\sourceg^\dagger\hat\phi_\sourceg^{\,}}}.$$
Classical experience with algorithms that generate the response for an arbitrary stimulus in a signal analysis formalism much more suggests nonlinearity than the \emph{a priori} linearity of the Wightman axioms and other mathematics for quantum fields.

\section{Nonlinearity in a real-space renormalization perspective}\label{RealSpaceRenormalization}
To make contact with real-space renormalization\cite[\S 4.4, \S 5.5]{KadanoffSHPMP} (or, in textbook form, \cite[Ch. 5]{BinneyEtAl}, for example) in an axiomatic test function approach, we suppose $\hat\xi_f$ is a quantum field measurement operator, with $f$ a test function on a square region that we split into $N$ fragments and $f^{[N]}_j\,{=}\,j$th of $N$ fragments of $f=\sum_{j=1}^N f^{[N]}_j$, then we apply the majority rule blocking algorithm (or some other),\\[0.5ex]
{\small\begin{tabular}{|c|c|c|}
\hline
$-1$ & $+1$ & $-1$\\
\hline
$+1$ & $+1$ & $-1$\\
\hline
$-1$ & $-1$ & $+1$\\
\hline
\end{tabular}}
$\ \longrightarrow -1$, for example, with $N=9$, effectively constructing $\hat\xi^{[N]}_f\,{=}\,\epsilon\!\left[\sum_{j=1}^N\epsilon\bigl[\hat\xi_{f^{[N]}_j}\bigr]\right]$.
We can iterate this\\[0.5ex] to construct $\hat\xi^{[N_1,...,N_k]}_f\,{=}\,\epsilon\Bigl[\sum_{j=1}^{N_k}\epsilon\bigl[\hat\xi^{[N_1,...,N_{k-1]}}_{f^{[N_k]}_j}\bigr]\Bigr]$.
Repeated blocking can be systematized in other ways, but there \emph{will} be nonlinearity, because otherwise $\hat\xi^{[{\cdots}]}_f$ would be no different from $\hat\xi^{\ }_f=\sum_{j=1}^N\hat\xi_{f^{[N]}_j}$.

Even though $\hat\xi^{[{\cdots}]}_f$ is nonlinear in $f$, it is arguably still a ``quantum field'', so that, in particular, it satisfies microcausality, $[\hat\xi^{[{\cdots}]}_f,\hat\xi^{[{\cdots}]}_g]\,{=}\,0$ if $f$ and $g$ are space-like separated.
There is no strict necessity that the fragments used must not overlap, nor that they must add up to the original test function, although either of those or other constraints might be empirically effective constraints.
We only require that the supports of all the fragments are contained in the support of the original test function.
In this form, we can say that we are just using the free field $\hat\xi_f$, or a subalgebra of it, however if we consider an infinite limit of progressive blocking (and rescaling) we can be sure only that $\hat\xi^{[{\cdots}]}_f$ must be nonlinear and it may not be constructible in any simple way from a linear theory, which we will take to be enough for the restricted aim of this Appendix: to justify weakening the Wightman axioms so that $\hat\xi_f$ is not a linear functional of the test functions, in an ad-hoc real-space renormalization argument that is intended only to support the argument in Section \ref{Nonlinearity}.

\end{document}